\documentstyle[prb,aps,multicol,epsfig,amsmath,amsfonts,psfrag,rotating]
{revtex}
   
 \newcommand{\nts}[1]{\tmspace{-}{#1\thinmuskip}{#1\txtmu}}
 \newcommand\eqlabel[1]{\label{#1}}
\allowdisplaybreaks
\begin{document}
\draft

\title{Ground State Energy Calculations of the 
\boldmath $ \nu=1/2 $ \unboldmath and the \boldmath $ \nu=5/2 $
\unboldmath FQHE System} 

\author{J. Dietel}
\address{
Institut f\"ur Theoretische Physik, Universit\"at Leipzig,\\
Augustusplatz 10, D 04109 Leipzig, Germany }
\date{\today}
\maketitle

\begin{abstract}
We reconsider energy calculations of the spin polarized
$ \nu=1/2 $ Chern-Simons theory.
We show that one has to be careful in the definition of the Chern-Simons
path integral in order to avoid an IR divergent magnetic ground state energy  
in RPA as in \cite{di1}. We correct the path integral and
get a well behaved magnetic
energy by considering the energy of the maximal divergent graphs
as well as the Hartree-Fock graphs. Furthermore, we consider the $\nu=1/2 $
and the $\nu=5/2$ system
with spin degrees of freedom. In doing this we formulate a Chern-Simons theory
of the $ \nu=5/2 $ system by transforming the interaction operator
to the next lower Landau level. We calculate the Coulomb energy
of
the spin polarized as well as
the spin unpolarized $\nu=1/2 $ and the $\nu=5/2$ system as a function of the
interaction strength in RPA.
These energies are in good agreement with numerical simulations of interacting electrons
in the first as well as in the  second Landau level. Furthermore, we calculate
the compressibility, the effective mass and the excitations of the
spin polarized $ \nu=2+1/\tilde{\phi} $ systems where $ \tilde{\phi}$ is an
even number.
\end{abstract}

\pacs{71.10.Pm, 73.40.Hm 71.27.+a}

\begin{multicols}{2}
\narrowtext

\section{Introduction \label{k1}}
The combination of an electronic interaction and a strong magnetic 
field in a two-dimensional electron system yields a rich variety of 
phases. These are best classified by the filling factor $ \nu $, 
which is the electron density divided by the density of a completely 
filled Landau level.
In this work we mainly consider energy calculations on   
systems with filling factors $\nu=1/2$ and $ \nu=5/2 $.
These system are most suitably described
by the Chern-Simons theory. In solid state physics this theory
is mainly used in the fields of the 
fractional quantum Hall effect and high temperature superconductivity.
The applications of the Chern-Simons theory in the field of 
high-temperature superconductivity is based on a work of Polyakov 
\cite{po1}.
Since the discovery of the fractional quantum Hall effect 
by Tsui, St\"ormer and Gossard (1982) \cite{ts1} there were many attempts to  
explain this experimental observation.
The current Chern-Simons type theories  of this 
effect are  mainly based on a work of Jain (1989) \cite{ja1}.    
In his theory he mapped the wave functions of the integer quantum Hall effect
to wave functions of the fractional quantum Hall effect. In the case of
filling fraction $ \nu=1/2 $ every electron gets two magnetic flux quantums
through  this mapping. By this transformation one gets new quasi-particles
(composite fermions) which do not see any
magnetic field in first approximation (mean field).
A field theoretical language for this scenario was first established by
Halperin, Lee, Read (HLR) (1992) \cite{hlr} as well as Kalmeyer and Zhang
(1992) \cite{ka1} for the $ \nu=1/2 $ system. The interpretation of
many experiments supports
this composite fermion picture. We mention transport experiments with
quantum (anti-) dots \cite{kan1}, and focusing experiments \cite{sm1} here.
An overview of further experiments can be found in \cite{wil2}.\\
HLR studied many physical quantities within the
random-phase approximation (RPA). Most prominent among these is the effective
mass of the composite fermions which they found to diverge at the Fermi
surface \cite{hlr,st1}. Recently, Shankar and Murthy \cite{sh1} proposed a
new theory of the $ \nu=1/2 $ system. Based upon a transformation
of the Chern-Simons Hamiltonian one achieves a separation
of the magneto-plasmon oscillators from the total interaction of the
system. After restricting the number of the magneto-plasmon oscillators
to the number of electrons they got a finite
quasi-particle mass which scales with the inverse of the strength of the
Coulomb repulsion. In this derivation they calculate a smaller number of
self energy Feynman graphs than in the RPA. Just recently
Stern et al. \cite{st2} calculated the self energy Feynman graphs 
for the theory of Shankar and Murthy in RPA and got
the same divergence of the
effective mass as HLR. Besides the theory of Shankar and Murthy there are
other alternate formulations of the composite fermionic picture which are
formulated entirely in the lowest Landau Level
\cite{pas1,ree1,le1}.\\
In the following we will use mainly the Chern-Simons theory of HLR to
calculate ground state energies of the $ \nu=1/2 $ and the $ \nu=5/2 $ system.
To get composite fermions one has to transform the electronic
wave function with a rather singular transformation. Due to this
transformation one gets 
a density-density interaction of the form
$ \sim 1/k^2 $ for small
wave vectors. Because of this singular interaction the singular diagrams in
leading order should be
resummed. These diagrams are well known as the RPA in the jellium problem
in $ d=3$. 
We have shown in an earlier publication 
\cite{di1} that despite of resumming these singular diagrams
one gets
an IR-divergence in the canonical potential of the $ \nu=1/2 $ system
in RPA due to the $ \sim 1/k^2 $ interaction.
Thereby we used the
Chern-Simons path integral of HLR for the resummation of the diagrams.
We obtained further that the Coulomb part of the energy is finite 
and in good agreement
with the Coulomb energy of numerical simulations of
interacting electrons in the lowest Landau level by Morf and d'Ambrumenil
\cite{mo1} as well as Girlich \cite{gi1}. We have to mention that 
C. Conti and T. Chakraborty \cite{co1} got an excellent agreement with the
numerical results by calculating the Coulomb energy 
with the help of the STLS theory \cite{sin1} which is a generalisation
of the RPA theory. 
In the following it is shown that we obtained an IR 
divergence of the magnetic energy in our earlier work
because the normal ordering of the
Chern-Simons Hamiltonian was not properly taken into account in the
path integral of HLR. In \cite{ap1} we correct this error with the help of an
intermediate time technique in the path integral and a change of the
coupling between the fermion fields and the bosonic fields. One can
show that this path integral has a finite energy in the RPA.
Details
of the analysis of the RPA for this path integral will be
published elsewhere.
In the case of the Coulomb gas, the Hartree-Fock graphs belong to the maximal divergent
Feynman graphs \cite{ge1}. This is no longer valid for the Chern-Simons theory.
For this theory we will get a finite energy for
the spin polarized $ \nu=1/2 $ system 
by calculating the energy of all Feynman-graphs which belong either to
the Hartree-Fock graphs or to the maximal divergent graphs (for a given
number of interaction vertices). With the
help of this principle one gets a well-suited finite
approximation of the magnetic energy.
Furthermore, we get the same Coulomb energy as in \cite{di1}. \\
In the second part of this work we will calculate the Coulomb energy of the
$ \nu=1/2 $ and the $ \nu=5/2 $ system
including the spin degree of freedom.
The $ \nu=5/2 $ system is of theoretical interest because it consists of
one Landau level filled with spin up and spin down electrons.
Like in the $ \nu=1/2 $ system the second Landau level is half filled with
electrons. So the $ \nu=5/2$   system should have similar physical properties
as the $ \nu=1/2 $ system. As a matter of fact this is not the case.
Eisenstein et al.\cite{ei1} were able to show with the help of tilted field experiments
that the ground state of the $ \nu=5/2 $ system is
spin unpolarized and incompressible.
This is different to the ground state of the $ \nu=1/2 $ system,
which is spin polarized and compressible.
Numerical simulations of the ground state of the $ \nu=1/2 $ and
the $ \nu=5/2 $ system  \cite{be1,mo2} 
show that one gets a transition from a spin unpolarized ground state to a
spin polarized  ground state depending on the strength of the interaction
between the electrons.
For calculating this transition with perturbational methods we will
construct a Chern-Simons theory of the
$ \nu=2+1/\tilde{\phi} $ systems where $ \tilde{\phi} $ is an even number
by transforming 
the Coulomb interaction of the second Landau level to the first Landau level.
By neglecting the lowest Landau
level, the $ \nu=5/2 $ system
behaves like a $ \nu=1/2 $ system with an altered
interaction potential.
Because of this we can calculate the Coulomb part of the energy of
the $ \nu=5/2 $
system within the RPA formalism
of the $ \nu=1/2 $ system. It will be shown that one gets a transition from a spin polarized to a spin unpolarized
ground state for the $ \nu=1/2 $ system as well
as for the
$\nu=5/2 $ system depending on the interaction strength. This transition
is in qualitative agreement
with the numerical simulations \cite{be1,mo2}. Furthermore, we will use
the Chern-Simons theory of the $ \nu=2+1/\tilde{\phi} $ systems to calculate 
the compressibility, the effective mass and the excitations. \\
As mentioned above we will treat the energy problem of the $ \nu=1/2 $ and
the $ \nu=5/2 $ Chern-Simons system in RPA. The use of this approximation
is motivated by the similarity of the interaction potentials of these two
systems to the one of the $ d=3 $ jellium problem. It is well known that for 
the latter
a perturbational calculation  of the
ground state energy results in a
good approximation especially for small densities
(which is a small parameter in this theory). Such a small parameter is not
existent in the $ \nu=1/2 $ Chern-Simons theory. Until now it is not clear
wether the RPA calculations of the ground state energy in
Chern-Simons theories are in 
agreement with the experiments. One aim of this paper is to make
one step further to a positive answer to this question. \\[0.3cm]  
The paper is organized as follows:
In section II  we reconsider the polarized $ \nu=1/2 $ system with the
help of a path integral which takes the normal ordering
of the Chern-Simons Hamiltonian into account. We calculate the energy of the
maximal divergent graphs together with the Hartree-Fock Feynman graphs of the
Chern-Simons Hamiltonian. In section III we formulate the RPA theory of the $
\nu=1/2 $ Chern-Simons theory subjected to a spin constraint. In
section IV we formulate a Chern-Simons theory of the
$ \nu=2+1/\tilde{\phi} $ systems  
and calculate the compressibility, the effective mass and the excitations
in RPA.
In section V we calculate the Coulomb ground state energy of
the spin polarized as well as 
the spin unpolarized $ \nu=1/2 $ and $ \nu=5/2 $ systems.
\section{The ground state energy of the spin polarized
  \boldmath$ \nu=1/2 $ \unboldmath system \label{k2}}
In this section 
we consider interacting spin polarized electrons moving in two dimension in a
strong magnetic field $ B $ directed in the positive $ z $-direction
of the system.
The electronic density of the system is chosen such that the lowest Landau
level of a non-interacting system is filled to a fraction
$ \nu=1/ \tilde{\phi}$ where $ \tilde{\phi} $ is an even number. We are mainly
interested in $ \tilde{\phi}=2 $. After performing a Chern-Simons
transformation \cite{zh2} of the electronic wave function one gets the
Hamiltonian of the composite fermions as:
\begin{eqnarray}
& & H_{CS}=\int d^2r
\bigg[\frac{1}{2m}\big|\big(-i\vec{\nabla}+\vec{A}  
+ \vec{a}_{CS}\big)\Psi(\vec{r})\big|^2 \eqlabel{1050}   \\
& &   +\frac{1}{2} \int d^2r' \Big\{
 (|\Psi(\vec{r})|^2-\rho_B)
 V^{ee}(|\vec{r}-\vec{r}\,'|)(|\Psi(\vec{r}\,')|^2-\rho_B)\Big\}\bigg].
 \nonumber 
\end{eqnarray}
The Chern-Simons operator $ \vec{a}_{CS} $ is defined
by $ \vec{a}_{CS}(\vec{r})=\tilde{\phi} \int d^2r' \;\vec{f}(\vec{r}-\vec{r}\,')
\Psi^+(\vec{r}\,')\Psi(\vec{r}\,') $. 
Here $ \Psi^+(\vec{r}) $ creates (and $ \Psi(\vec{r}) $ annihilates) a
composite fermion
with coordinate $ \vec{r} $. $ V^{ee}(r)=e^2/r $ is the Coulomb interaction
where $ e^2=q_e^2 /\epsilon  $. $ q_e $ is the charge of the electrons and
$ \epsilon $ is the dielectric constant of the background field $ \rho_B $.
$ \vec{A}(\vec{r}) $ is the vector potential \
$ \vec{A}=1/2 \,\vec{B} \times \vec{r} $ and $\vec{B} $ is a 
homogeneous magnetic field in z-direction $ \vec{B}=B \vec{e}_z $
where $ \vec{e}_z $ is the
unit vector in $z $-direction. 
We suppose throughout this paper that $ B $ is a positive number. 
The function $ \vec{f}(\vec{r}) $ is given by 
$ \vec{f}(\vec{r})=-\vec{e}_z \times \vec{r}/r^2 $.
We used 
the convention $ \hbar=1 $ and $ c=1$ in the above formula
(\ref{1050}). Furthermore, we set $ q_e=1 $ for the coupling of the magnetic
potential to the electrons. It is well known, that the partition function of
the Hamiltonian (\ref{1050}) can be written in an operator formalism
\cite{po1} with the help of
the bosonic Chern-Simons fields ($ a^0(r,t), \vec{a}(r,t) $).
This is shown in the first
quantized path integral language.
Using standard methods \cite{ne1} we can transform
this partition function to a path integral. This path integral is written  
as 
\begin{eqnarray}
Z_{1/2}& = &  \lim_{\epsilon \to 0} \frac{1}{\overline{N_{1/2}}}
\prod\limits_{l=1}^{N_l}
\int_{\tiny \mbox{BC}} {\cal D}
[a^0_l,\vec{a}_l,\sigma_l]\, {\cal D}[\Psi^*_l,\Psi_l] \eqlabel{1150} \\
& & \times 
\exp\left[-\epsilon\left( 
L_{l}+L_{CS,l}+L_{ee,l}+L^0_{ee,l}\right)\right] \nonumber \;. 
\end{eqnarray}
The various functions in (\ref{1150}) are given by 
\begin{eqnarray}
  L_{l} & = &    \int d^2r\; \Psi^*_{l}(\vec{r})
 \frac{1}{\epsilon} \left(\Psi_l(\vec{r})- \Psi_{l-1}(\vec{r})\right)
 \eqlabel{1160}\\
& &   -\Psi^*_{l}(\vec{r}) \left( \mu+\left(1+
    \frac{\epsilon}{2}a^0_{l}(\vec{r})\right)a^0_{l}(\vec{r})\right)
 \Psi_{l-1}(\vec{r})\nonumber\\
& &   + \frac{1}{2m}\Psi^*_{l} (\vec{r})
\left(-i\vec{\nabla}+\vec{A}(\vec{r})+\vec{a}_l(\vec{r})
\right)^2\Psi_{l-1}(\vec{r}) \;,\nonumber  \\
\nts{70} L_{CS,l} & = &   
\frac{1}{2\pi\tilde{\phi}}\int
d^2r\;a^0_{l}(\vec{r})\,\vec{\nabla} \times \,\vec{a}_{l}(\vec{r}) 
\eqlabel{1165} \;,\\  
L_{ee,l} & = &   
\int d^2r\,d^2r'
\sigma_{l}(\vec{r}) \;V^{ee}(|\vec{r}-\vec{r}\,'|) \nonumber  \\
& & \hspace{2cm} \times (\Psi^{*}_{l}(\vec{r}\,')
  \Psi_{l-1}(\vec{r}\,')-\rho_B)\;,  \eqlabel{1170}   \\
L^0_{ee,l} & = & -\frac{1}{2}\int
d^2r\,d^2r\,' \;
\sigma_{l}(\vec{r})\;V^{ee}(|\vec{r}-\vec{r}\,'|) \;
\sigma_{l}(\vec{r}\,') \;,\eqlabel{1167} 
\end{eqnarray}
together with the norm
\begin{equation}
\overline{N}_{1/2}=\prod\limits_{l=1}^{N_l}\int_{\tiny \mbox{BC}}{\cal D}[a^0_l,\vec{a}_l,\sigma_l]
\exp\left[-\epsilon\left(L_{CS,l}+L^0_{ee,l}\right)\right]      \;.\eqlabel{1175}
\end{equation}
The path integral (\ref{1150}) is correct under the  
gauge condition $ \vec{\nabla} \cdot \vec{a}_l=0 $.
$ a^0_l $ is a imaginary field. $ \vec{a}_l $ and $ \sigma_l $ are real
fields. 
The time slices $ \epsilon $ are defined as $\epsilon=\beta/N_l $.  
The index $l $ counts the discrete time slices. 
Furthermore, we have anti-periodic boundary conditions $\Psi_{N_l}=-\Psi_{0}$
for the Grassmann fields \cite{ne1}.
The action of the path integral (\ref{1150}) is given by a fermionic term $
L_{l} $, a bosonic term $ L_{CS} $ of the Chern-Simons form, and two 
Coulomb interaction terms $ L_{ee,l}$, $ L_{ee,l}^0 $. 
In 
comparison to the Chern-Simons path integral  
of HLR \cite{hlr} which was used in our earlier publication \cite{di1}
we get an additional term proportional 
to $ \epsilon a^0_l(\vec{r})^2 $  in $ L_l $ (\ref{1160}). 
This term is mainly due to the normal-ordering 
of the $ \Psi^6 $ term in the Chern-Simons Hamiltonian $ H_{CS} $ 
(\ref{1050}). This is most easily seen by integrating the path integral 
(\ref{1150}) over the Chern-Simons fields. Due to the additive term 
one can not perform the formal limit $\epsilon \to 0 $ in (\ref{1150}). 
This limit has to be taken in every additive term after integration over  
the Chern-Simons and fermionic fields. Furthermore, we have 
to remark, that we treat the same formalism as above in order to
derive 
the path integral of Shankar and Murthy \cite{sh1,st2}.
There we get no additional term in comparison to their 
path integral \cite{di3}.\\
In the following we integrate (\ref{1150}) over the fermionic fields.
Using a mean field expansion of the result up to the second order
in the bosonic
fields one gets after integration for the
grand-canonical potential $ \Omega= \Omega_0+ \Omega_{\tiny \mbox{RPA}} $.
$ \Omega_0 $ is the grand-canonical potential of the Coulomb-free
electron gas.
We now split $ \Omega_{\tiny \mbox{RPA}} $ into a variety of terms
for which one
can take the limit $ \epsilon \to 0 $ rather easily \cite{di1}.
\begin{equation}
\Omega_{\mbox{\tiny RPA}}=\Omega_R+\mbox{F}^{ee}+\mbox{F}^{CS}+
\mbox{H}  \;. 
\end{equation}
The expression $ \Omega_R $ corresponds to the RPA graphs which do not
belong to the Hartree-Fock graphs. $ \mbox{F}^{ee}+\mbox{F}^{CS} $ are the
Coulomb and Chern-Simons Fock diagrams. $ \mbox{H} $ are the Hartree
diagrams. We will see below that the Hartree term  is due to the normal
ordering of the $ \Psi^6 $-term in (\ref{1050}). 
For temperature $ T=0 $, $ \Omega_R $ 
is given by      
\begin{eqnarray} \eqlabel{5105}
\nts{4}\Omega_{R}& = &
\frac{1}{2(2\pi)^3} \int d\omega \,d^2q 
\;\Big[\log\left(1+\mbox{CS}(q,\omega)+\mbox{EM}(q,\omega)\right)  \nonumber \\
& & \hspace{3cm}
-\mbox{CS}_1(q,\omega)-\mbox{EM}(q,\omega) \Big] \;.
\end{eqnarray}
The functions $ \mbox{CS} $, $\mbox{CS}_1 $ and $ EM $ are defined as
\begin{eqnarray} 
\mbox{CS} & = & -(2\pi \tilde{\phi})^2 \frac{1}{q^2}
 \Pi_{a^0 a^0}
 \left( \Pi_{a a} 
 +\frac{\mu}{2\pi}\right)  \;, \eqlabel{5110}\\
\mbox{CS}_1 & = & -(2\pi \tilde{\phi})^2 \frac{1}{q^2}
 \Pi_{a^0 a^0} \frac{\mu}{2\pi}
     \;, \eqlabel{5120} \\
\mbox{EM}& = & - e^2\frac{2\pi}{q}
 \Pi_{a^0 a^0 }\,. \eqlabel{5130}
\end{eqnarray}

$ EM $ is the Coulomb term and $ CS $ is the Chern-Simons term of the
RPA formula. The second and third  summands  $ EM $ and $ CS_1 $ in
(\ref{5105}) are  given by the first order graphs of the RPA
which have to be subtracted
for treating equal time Green's function
in the right way \cite{di1}. 
The density-density response function 
$ \Pi_{a^0 a^0}\left(q,\omega,\mu\right) $ and the transversal
current-current response function  
$ \Pi_{a a}\left(q,\omega,\mu\right) $ are given in (\ref{30280})
of appendix A. 
The Hartree-Fock terms  of $ \Omega_{\tiny \mbox{RPA}} $ are  given by 
\begin{eqnarray}
\mbox{H} & = & \frac{1}{2} (2\pi \tilde{\phi})^2 \frac{m \mu^2}{(2\pi)^4} 
\int d^2q \,\frac{1}{q^2} \;,\eqlabel{5135} \\ 
 \mbox{F}^{XX} & = & \frac{1}{2\,(2\pi)^3},\eqlabel{5140}  \\
 && \nts{10} \times \int d^2k d^2q \,n_F(k,\mu)\,
 n_F(q,\mu)\,V^{XX}(|\vec{k}-\vec{q}|)  \nonumber 
 \end{eqnarray}
The function $  n_F(q,\mu) $ is the fermion occupation factor for the chemical
potential $\mu $. 
 The vertices $ V^{XX} $ are defined by
\begin{equation}\eqlabel{5145}
 V^{ee}(q)=e^2\frac{1}{q} \quad , \qquad 
 V^{CS}(q)=\tilde{\phi}^2 \mu \frac{1}{q^2}  \;.
\end{equation}

Here $ V^{ee} $ is the Coulomb vertex. $ V^{CS} $ is the
Chern-Simons vertex $ \sim 1/q^2 $ mentioned in the introduction. 
Comparing $ \Omega_{\tiny\mbox{RPA}} $ with the analogous expression in
\cite{di1} the only difference between the two
expressions is the  Hartree term $ \mbox{H} $ in (\ref{5135}),
which corresponds to a
Hartree Feynman graph of $ H_{CS} $ (\ref{1050}) resulting from the normal
ordering of the $ \Psi^6$ term.
It is shown in \cite{di1}, that $ \Omega _{\tiny\mbox{RPA}}-\mbox{NO} $ is
IR-divergent. This  is based on the IR-divergence 
in $\mbox{F}^{CS} $.
The additive term $ \mbox{H} $ shifts this IR-divergent grand canonical
potential
in \cite{di1} to an UV-divergent grand
canonical potential $ \Omega_{\tiny\mbox{RPA}} $. 
As mentioned above $\mbox{F}^{CS} $ and
$ \mbox{H} $ belong to the Hartree-Fock Feynman graphs of
$H_{CS} $. Therefore, if we calculate the energy
of all Hartree-Fock Feynman graphs in addition to $ \Omega_R $ we
will get a finite energy because the
Hartree-Fock energy of $ H_{CS} $ (\ref{1050}) has to be finite.
Furthermore, it is easily seen that
$ \Omega_R $ corresponds to the maximal divergent graphs for a
given number of vertices. 
To get a good approximation of the energy we take into account in the following
the maximal divergent graphs (per
number of vertices) together with the Hartree-Fock graphs.
As mentioned in the introduction this energy principle is also
used for a calculation of the ground state energy of the Coulomb gas by many
authors (e.g. \cite{ge1}).   
In order to calculate that energy we have to determine at first
the Hartree-Fock energy of $ H_{CS} $. 
Sitko calculated in \cite{sit1} the Hartree-Fock energy of the
magnetic part of $ H_{CS} $. The Coulomb part of the
Hartree-Fock energy consists
only of  $ \mbox{F}^{ee} $ (\ref{5140}). This integral is easily evaluated
yielding 
\begin{equation} \eqlabel{2315}
U_{HF}=\frac{m}{4 \pi} \mu^2+\frac{3 m }{16 \pi} \tilde{\phi}^2 \mu^2
- \frac{2 \sqrt{2}}{3 \pi^2} \, e^2 m^{\frac{3}{2}}
 \mu^{\frac{3}{2}}
\end{equation}
for the Hartree-Fock energy of the Chern-Simons Hamiltonian
$ H_{CS}$. 
The first term in (\ref{2315}) is the kinetic energy of the
composite fermions.
For the $ \nu=1/2 $ system we have to insert $ \tilde{\phi}=2 $ into
$ U_{HF} $. Doing this we 
obtain two times the value of the exact magnetic energy
for $ U_{HF} $ ($ \mu=(2 \pi) \rho /m $). \\
With the help of an $ e^2 $-expansion of $ \Omega_R $ in (\ref{5105})
we can calculate the coefficients of the expansion for the $
\nu=1/2 $ system numerically:
\begin{equation}
  \Omega_R=-0.19 \,m\, \mu^2 -0.038 \,e^2 m^{\frac{3}{2}}\mu^{\frac{3}{2}}+ O(e^4)  \eqlabel{2320} \;.
\end{equation}
The elctron density $ \rho $ is given by $ \mu=\rho/(2 \pi m) $.
This can be motivated by the
Luttinger-Ward theorem \cite{lu1} under the consideration that there are
no anomalous graphs in our approximation of the energy. 
Thus we get from
(\ref{2315}), (\ref{2320}) 
\[
U^{\mbox{\tiny Th}}\approx  5.06 \frac{\rho^2}{m}-2.11 e^2 
\rho^{\frac{3}{2}}\, ,\,
U^{\mbox{\tiny Num}} \approx  6.28 \frac{\rho^2}{m}-1.67 e^2
\rho^{\frac{3}{2}}
\]
for the energy density
$ U^{\mbox{\tiny Th}}=U_{HF}+ \Omega_R$ 
of the $ \nu=1/2 $ system.      
It is evident by the selection principle of the calculated Feynman graphs  
that the Coulomb
energy part in
$ U^{\mbox{\tiny Th}}$ is the same
as in \cite{di1}.
To  compare our perturbative result for the energy with the energy
obtained by other methods
we have written 
down in $ U^{\mbox{\tiny Num}}$ the exact magnetic
energy and the Coulomb energy obtained by numerical diagonalization
techniques.
This Coulomb energy was calculated by Morf and d'Ambrumenil
\cite{mo1} as well as Girlich \cite{gi1} by 
diagonalization of the Coulomb part of the Hamiltonian
for electrons on a sphere in the lowest Landau
level. One sees 
that perturbational and the numerical calculated energies are in satisfactory
agreement.

\section{The Coulomb energy of the \boldmath$\nu=1/2 $ \unboldmath system
 including constraints on the total spin \label{k3}}

In the $ \nu=1/2 $ as in the $ \nu=5/2 $ system appears an interesting phase
transition from a spin unpolarized ground state to a
spin polarized ground state by modifying the interaction strength
between the electrons. In the next few sections we will see, that this
physical phenomenon already exists on the level of the RPA
of the Chern-Simons theory. To show this we have to add a spin constraint
variable into the Chern-Simons path integral. Without this spin-constraint
variable the $ \nu=1/2 $ Chern-Simons path integral is written as in
(\ref{1150}) with a doubling of the fermionic Grassmann fields representing
the spin-up and spin-down freedom of the electrons.
 In the following we denote by $c^{+}_{\uparrow} (\vec{r})$,
 ($c^{+}_{\downarrow} (\vec{r})$) the creation operator of one composite fermion at
 point $ \vec{r}$  with
 spin up (spin down). $c_{\uparrow} (\vec{r})$,
($c_{\downarrow} (\vec{r})$) annihilates one composite fermion  with spin up
(spin down) at point $\vec{r} $.  \\
The total spin in the xy-plane is given by 
\begin{eqnarray}
S_x^2+S_y^2 & = &  \int d^2r 
\left(\frac{1}{2}c^{+}_{\uparrow}(\vec{r})\;c_{\uparrow}(\vec{r})+
\frac{1}{2}c^{+}_{\downarrow}(\vec{r})\;c_{\downarrow}(\vec{r})\right)+
S^2_{\mbox{\tiny Mod}}, \nonumber \\
S^2_{\mbox{\tiny Mod}} & = &  -\int d^2r\,d^2r'\;
c^{+}_{\uparrow}(\vec{r})\;c^{+}_{\downarrow}(\vec{r}\,')\;
c_{\downarrow}(\vec{r})\; c_{\uparrow}(\vec{r}\,')\,. \eqlabel{30110}
\end{eqnarray}
Furthermore, we define from $H_{CS}$ (\ref{1050})  
a new spin constraint Chern-Simons Hamiltonian $ H^s_{CS} $ by
$ H^s_{CS}=H_{CS}-\mu_s S^2_{\mbox{\tiny Mod}} $.
We will fix the total spin 
$ S^2=S_x^2+S_y^2+S_z^2 $ by the constraint  
such that $ S^2 $ is  minimal for a given  $ S_z $.
This is the case if
$ S^2\;= \;|S_z|\left(|S_z|+1\right) $. 
One can get this constraint on $ S^2 $ by differentiating  
$ -1/\beta \log(Z^s) $ with respect to $ \mu_s $.
Thereby we denote the partition function of $H^s_{CS}$ by $Z^s$. 
Thus we get 
the following conditions on the partition function $Z^s $  
\begin{eqnarray} 
& & -\frac{1}{\beta}\left(\frac{\partial}{\partial \mu_{\uparrow}}+
\frac{\partial}{\partial \mu_{\downarrow}}\right)\log(Z^s)=\langle|N|\rangle
\nonumber \\
& &  -\frac{1}{2}\frac{1}{\beta}
\left(\frac{\partial}{\partial \mu_{\uparrow}}
-\frac{\partial}{\partial \mu_{\downarrow}}\right)\log(Z^s)=\langle|S_z|\rangle
 \;, \nonumber  \\ 
& &  \frac{1}{\beta}\frac{\partial}{\partial \mu_s}\log(Z^s)  =
 -\frac{1}{4}\frac{1}{\beta^2}\left(\frac{\partial}{\partial
     \mu_\uparrow}-\frac{\partial}{\partial \mu_\downarrow}\right)^2 
\log(Z^s) \nonumber \\
& & \hspace{3cm}  -\frac{1}{2}
\langle |N|\rangle+\left|\langle|S_z|\rangle\right| \;. \eqlabel{30145}
\end{eqnarray}
The symbol $ \langle |\cdot |\rangle $
is the expectation value in the Gibb's state of 
the Hamiltonian $ H_{CS}^s $. $ N $ is the particle number operator 
$ N=\int d^2r\, c^+_\uparrow(\vec{r}) c_\uparrow (\vec{r})+
c^+_\downarrow(\vec{r}) c_\downarrow(\vec{r}) $.  
$ \mu_\uparrow $ and $\mu_\downarrow $ are 
the chemical potentials of the system representing the spin-up and
spin-down electrons.
Furthermore, we have to mention that one has to be careful
in using equation (\ref{30145}).
This is because the first term on the
right hand side of equation (\ref{30145}) 
equals to  $\langle|S_z^2|\rangle-
\langle|S_z|\rangle^2 $ for a finite system.
This is no longer the case for an infinite system (an example is the kinetic
Hamiltonian). 
Thus one has to differentiate first
for a finite system and afterwards take the limit to an infinite system.  
We are now able to write down the $ \nu=1/2 $ Chern-Simons path integral.
Using the Hubbard-Stratonovich decoupling of the fermionic fields in
the operator  
$ S^2_{\mbox{\tiny Mod}} $ we get 

\begin{eqnarray} 
& & Z_{1/2}^s= \lim_{\epsilon \to 0} 
\frac{1}{\overline{N}^s_{1/2}}\prod\limits_{l=1}^{N_l}
\int_{\tiny \mbox{BC}}{\cal D}[a^0_l,\vec{a}_l,\sigma_{l},\sigma^s_{l}]
{\cal D}[c^*_{\kappa,l},c_{\kappa,l}] \eqlabel{30150}  \\
 & & \hspace{0.2cm}\times \exp\left[-\epsilon\left( 
L^s_{l}+L_{s,l}+L_{ee,l}+L^0_{ee,l}
+L_{CS,l}+L^0_{s,l}\right)\right]  \nonumber 
\end{eqnarray}
with the help of 
\begin{eqnarray}
L^s_{l}& = &  \sum\limits_{\kappa \in \{\uparrow,
  \downarrow\}} \nts{1}\int d^2r\;
c^*_{\kappa,l}(\vec{r})\frac{1}{\epsilon}
\left(c_{\kappa,l}(\vec{r})-c_{\kappa,l-1}(\vec{r})\right) \nonumber \\
& &\nts{5} - c^*_{\kappa,l}\left(
\mu_\kappa+\left(1+\frac{\epsilon}{2}a^0_l(\vec{r})\right)a^0_l(\vec{r})\right)
 c_{\kappa,l-1}(\vec{r}) \nonumber  \\
& & \nts{5}+ c^*_{\kappa,l}(\vec{r})\frac{1}{2m}
\left(-i\vec{\nabla}+\vec{A}(\vec{r})+\vec{a}_l(\vec{r})
\right)^2 c_{\kappa,l-1}(\vec{r}),\eqlabel{30152}\\
L_{ee,l} & = & 
\int \; d^2r\,d^2r\,' \;
\sigma_{l}(\vec{r}) \;V^{ee}(|\vec{r}-\vec{r}\,'|)  \nonumber \\
& & \hspace{1cm} \times \bigg(
\sum\limits_{\kappa \in \{ \uparrow, \downarrow \}}
  c^{*}_{\kappa,l}(\vec{r}\,') c_{\kappa,l-1}(\vec{r}\,')-\rho_B\bigg) \;, \eqlabel{30154} \\
L_{s,l} & = & \mu_s \int
d^2r\,d^2r'\;
\sigma^s_l(\vec{r}) \left( 
c^*_{\uparrow,l}(\vec{r}\,')\;c_{\downarrow,l-1}(\vec{r}\,')\right) \;,
\eqlabel{30156}\\
 L^0_{ee,l}& = & -\frac{1}{2}\int
d^2r\,d^2r' \;
\sigma_{l}(\vec{r})\;V^{ee}(|\vec{r}-\vec{r}\,'|) \;
\sigma_{l}(\vec{r}\,') \;,\eqlabel{30158} \\ 
L^0_{s,l} & = & -\frac{\mu_s}{4} \int d^2r\,d^2r'\;\left(
\sigma^s_{l}(\vec{r}) \sigma^s_{l}(\vec{r}\,')\right)  \eqlabel{30160}
\end{eqnarray}
and the normalizing factor 
\begin{eqnarray}
\overline{N}^s_{1/2} & = & \prod\limits_{l=1}^{N_l}\int_{\tiny \mbox{BC}}
{\cal D}[a^0_l,\vec{a}_l,\sigma_{l},\sigma^s_{l}] \nonumber \\
 & & \hspace{1cm} \times
 \exp\left[-\epsilon\left(L_{CS,l}+L^0_{ee,l}+L^0_{s,l}\right)\right]
\eqlabel{30170}\;.
\end{eqnarray}

The terms $ L_{s,l} $, $L^0_{s,l} $ in (\ref{30150}) are given 
by the Hubbard-Stratonovich decoupling of the  spin term 
$ S^2_{\mbox{\tiny Mod}} $. 
After integrating in (\ref{30150}) over the fermionic fields we get
the following mean-field equations for the bosonic fields 
\begin{equation} \eqlabel{30210}
\begin{array} {c c} 
\vec{\nabla} \times \vec{a}_l=(2 \pi \tilde{\phi})\, B 
&\quad , \qquad   a^0_l=0  \;, \\
\sigma_l=0    &  \quad , \qquad\sigma^s_l=0 \;.
\end{array}
\end{equation}
The mean-field Green's function is given by $ G^s(q,\omega)\; =\;(
    -1/(i\omega-q^2/(2m)+\mu_\uparrow),    
-1/(i\omega-q^2/(2m)+\mu_\downarrow))$.
Integrating this mean-field expansion to second order of the bosonic
fields we get for the grand canonical potential
$ \Omega^s_{1/2}= \Omega^s_0+\Omega^s_{\tiny \mbox{RPA}} $.
$ \Omega^s_0 $ is the grand
canonical potential for a spin dependent electron gas without
interaction. As we do not have any anomalous graphs,
we want to make use of the Luttinger-Ward theorem \cite{lu1}. 
This theorem states that a good approximation is obtained for 
$ \mu_\uparrow $ , $ \mu_\downarrow $ if one uses 
 $m(\mu_\uparrow+\mu_\downarrow)/(2 \pi) =\rho $ and
$ m (\mu_\uparrow-\mu_\downarrow)/(4 \pi)= \langle|S_z|\rangle/F $.
$ F $ is the area of the system.
As a function of $ \mu_\uparrow $, $ \mu_\downarrow $, the expectation value 
$ \langle| S_z |\rangle $ behaves like a step function 
for a finite system.
Because the first term on the right hand side of equation (\ref{30145})
is  $ - 1/(2 \beta)\, (\partial/\partial
     \mu_\uparrow-\partial/\partial \mu_\downarrow) \,
\langle|S_z |\rangle $ we observe that this term is zero.  
Because $ \Omega_0 $ does not depend on $ \mu_s $, it can not
fulfill the spin constraint equation (\ref{30145}).
On the other hand it is clear that the ground state
of the mean-field Hamiltonian fulfills the spin constraint 
$\langle|S^2|\rangle=|\langle S_z\rangle |(|\langle S_z\rangle |+1) $.
For this reason $ \Omega_0 $ together with the grand canonical
potential of the
exchange graph of $ S^2_{\tiny \mbox{Mod}} $ (\ref{30110}) fulfills the spin constraint
equation (\ref{30145}) (The exchange graph is a part of $ \Omega^s_{\tiny\mbox{RPA}} $).
Furthermore, we have to mention that the exchange part of
$ S^2_{\tiny \mbox{Mod}} $ is the only 
part of $ \Omega^s_{\tiny\mbox{RPA}} $ which is linear in $ \mu_s $.
Thus the spin constraint equation
(\ref{30145}) is fulfilled for $ \mu_s=0$.
In the following we denote $ \Pi(q,\omega,\mu_\uparrow)+\Pi(q,\omega,\mu_\downarrow) $
by
$ \Pi^d(q,\omega,\mu_\uparrow,\mu_\downarrow) $. $ \Pi $  are the
diverse response functions of appendix A.  After some transformations
and an expansion in $e^2 $ we get with the help of the functions 
\begin{eqnarray}
\mbox{CS}^s & = & -(2\pi \tilde{\phi})^2 \frac{1}{q^2} \Pi^d_{a^0 a^0}
\left(\Pi^d_{a a}+\frac{\mu_{\uparrow}+
\mu_{\downarrow}}{2\pi}\right) \;, \eqlabel{30356}  \\  
\mbox{EM}_{1/2} & = & - (2 \pi) V^{ee}(q)\,\Pi^d_{a^0 a^0} \;,\eqlabel{30358}  \\
\mbox{F}_{1/2}^{ee} & = &-\nts{5}\sum\limits_{\kappa \in \{ \uparrow,\downarrow
  \}}\nts{2}\frac{1}{2(2\pi)^3}  \eqlabel{30360} \\
& & \times \int d^2k d^2q V^{ee}(|\vec{k}-\vec{q}|)n_F(k,\mu_\kappa)
n_F(q,\mu_\kappa) 
\nonumber 
\end{eqnarray}
the $ e^2 $-part of the grand-canonical $ \Omega^s_{1/2} $ as 
\begin{eqnarray}
 \Omega_{1/2}^{ee}&=& \mbox{F}_{1/2}^{ee} \eqlabel{30370}\\
& & \nts{20} +\frac{1}{2(2\pi)^3}\int
 d\omega \,d^2q \left(
\frac{1}{(1+\mbox{CS}^s(q,\omega))}-1\right)\mbox{EM}_{1/2}(q,\omega). \nonumber 
\end{eqnarray}
After developing the Chern-Simons theory of the $ \nu=5/2 $ system
in the next section, we will calculate $\Omega_{1/2}^{ee}$ in section V
numerically.

\section{The Chern-Simons theory of the \boldmath $ \nu=5/2 $
  \unboldmath system \label{k4}}
The $ \nu=5/2 $ system consists of one lowest Landau level filled by spin up
and spin down electrons. The second Landau level is half filled. After
performing 
the Chern-Simons transformation of the Hamiltonian of this system one does
not get a mean field free theory as in the case of the $ \nu=1/2 $ system.
It is clear that the filled lowest Landau level of the $ \nu=5/2 $ system
should not have much relevance for the physical properties.
Thus we will consider the $ \nu=5/2 $ system as a $ \nu=1/2 $ system with a
modified interaction potential between the electrons. If we define an
isometric transformation which transforms a wave function of a higher
Landau level to the next lower Landau level then the calculation of
the Coulomb energy in the second Landau level corresponds to the calculation
of the energy of the first Landau level with the transformed interaction
potential of the electrons. 
If one calculates the Coulomb energy by perturbational methods then
this transformation should not change the wave function too much.
We will develop in the following an isometric transformation which leaves
the center of mass of the wave function invariant.\\ 
The ladder operators between Landau levels are defined
with the help of the operators
$ \Pi_L^x \; = \;  -i\partial_x+A^x $ and 
$\Pi_L^y \; = \;-i\partial_y+A^y $ by
\begin{equation} \eqlabel{30010}
\Pi_L \;= \;\Pi_L^x-i\,\Pi_L^y \quad , \qquad
\Pi_L^{+} \; =\; \Pi_L^x+i\, \Pi_L^y \;.
\end{equation}
The ladder operators act on a wave function belonging to the n-th
Landau level as follows
\begin{eqnarray}\eqlabel{30030}
\Pi_L^{+} |n,q> & = &  \sqrt{2 B(n+1)}\, |n+1,q> \;, \nonumber \\
\Pi_L |n,q> & = &   \sqrt{2Bn}\, |n-1,q> 
\end{eqnarray}
With the help of some commutation relations \cite{sp1} one can easily see that  
\begin{eqnarray}
& & \frac{1}{2 B(n+1)}\nts{2} <n,q|\Pi_L\,
x \,\Pi_L^{+} |n,q>=  <n+1,q|x|n+1,q>  \nonumber \\
& & \frac{1}{2 B n}\nts{2} <n,q|\Pi_L^+ \, x \,\Pi_L|n,q> =   <n-1,q|x|n-1,q>
\eqlabel{30035} 
\end{eqnarray} 
The same relation holds for the $ y $-coordinate.
 The partial isometric transformation $ P $ which is descending 
the Landau level   
 functions is given by
\begin{equation}
  P |n,q>\;:= \; |n-1,q>  \;.\eqlabel{30038}
 \end{equation}
With the help of the ladder  
 operators (\ref{30010}) 
 the operator $ P $ is written by
\begin{eqnarray}
P & = & \frac{\Pi_L}{(2B)^\frac{1}{2}}+
\Pi_L^+\Pi_L^2\frac{1}{(2B)^\frac{3}{2}\;\sqrt{2}}
\left(1-\sqrt{2}\right)  \nonumber \\ 
& & +\Pi_L^{+2}\Pi_L^3\;\frac{1}{(2B)^\frac{5}{2}\,2\sqrt{3}}\left(1+\sqrt{3}
\left(1-\sqrt{2}\right)\right) \nonumber \\
& &+ \Pi_L^{+3}\Pi_L^4\,O\left(\frac{1}{B^{\frac{7}{2}}}\right)   \;.\eqlabel{30080}
\end{eqnarray}
It is easily seen from equation  
 (\ref{30080})  that the n-th term in $ P$ transforms a wave function 
which  belongs to the n-th lowest Landau level to zero.
The higher order terms in $ P$ are motivated by the different  
normalization factors of the individual Landau levels in equation
(\ref{30030}).
We observe from the equations (\ref{30030}) and (\ref{30035}) that the operator $ P$
leaves the center of mass coordinate of one Landau level function  
invariant. \\
We will use this operator $ P$ for transforming the Coulomb interaction
to one lower Landau level.
Almost all publications carrying out ground state energy calculations
of the $\nu=1/2 $ and the $ \nu=5/2 $ system
by numerical methods do not take into account  
Landau level splitting  \cite{mo1,mo2,be1}.  
In these calculations one diagonalizes the Coulomb interaction operator 
in the second Landau level in the case of the $ \nu=5/2 $ system. 
It is easily seen from perturbation theory that this energy is given 
by the $e^2 $-term of the Coulomb energy.
Because we want to compare the energy 
given by perturbational calculations and numerical  
calculations we can neglect the higher order $1/B $ terms in $ P$  
under the condition of calculating the Coulomb energy to order $ e^2$. 
So we get for the isometric partially 
transformed Coulomb interaction operator     
\begin{eqnarray}\eqlabel{30090}
{\cal V}_{5/2}^{ee}& = &
e^2 \frac{1}{(2B)^2} \int\,
d^2r\,d^2r' \; \Pi_L \Psi^{+}(\vec{r})\, \Pi_L \Psi^{+}(\vec{r}\,') \nonumber
\\ 
& & \hspace{1.5cm} \times \frac{1}{|\vec{r}-\vec{r}\,'|} \;
\Pi_L^+ \Psi(\vec{r}\,')\, \Pi_L^+\Psi(\vec{r}) \;.
\end{eqnarray}
$ B $ is given by $ (2 \pi \tilde{\phi}) \rho $ where
$ \rho $ is the density of the electrons in the second 
Landau level.
Since one does not start from a Chern-Simons transformed wave function  
in the lowest Landau level in the perturbational calculation 
but from a Slater determinant of plane waves, one should also take 
into account higher order terms of $ P$ in perturbation theory. 
According to equation (\ref{30030}) these only differ from the
first order term 
by a normalizing factor.
Therefore we will fix an effective $ B $-field 
$ B_{\tiny \mbox{eff}}$ in equation (\ref{30090}) when proceeding
\begin{equation}
 P \approx  (1/\sqrt{2B_{\tiny \mbox{eff} }})\,
\Pi_L \eqlabel{30091}
\end{equation}

This effective field will be brought into line with property (\ref{30080}).
The Chern-Simons transformed operator $ P_{CS} $ of $ P$ is
constructed from $ P$ in (\ref{30080}) by the substitution
$ \vec{A} \rightarrow \vec{A}+\vec{a} $. It is approximated through
$  P_{CS}\approx  (1/\sqrt{2B_{\tiny \mbox{eff} }})\,
\Pi_{L,CS}
$ 
with the operator $ \Pi_{L,CS} $ 
constructed from $\Pi_{L}$ by the substitution
$ \vec{A} \rightarrow \vec{A}+\vec{a} $. The $ \nu=5/2 $ Chern-Simons
path integral $ Z_{5/2}^s $
is constructed from the $ \nu=1/2 $ Chern-Simons path integral
$ Z_{1/2}^s $ by the substitution $ L_{ee,l}\rightarrow L^{5/2}_{ee,l} $.
The function $ L^{5/2}_{ee,l} $ is given by
\begin{eqnarray}
L^{5/2}_{ee,l}& = &   
\int \; d^2r\,d^2r' \;
\sigma_{l}(\vec{r})\;V^{ee}(|\vec{r}-\vec{r}\,'|) \eqlabel{30093}  \\ 
& & \times \bigg(\sum\limits_{\kappa \in \{\uparrow, \downarrow\}}
  P_{CS}\,c^{*}_{\kappa,l}(\vec{r}\,') \;
P^{+}_{CS} \,c_{\kappa,l-1}(\vec{r}\,')-\rho_B\bigg). \nonumber 
\end{eqnarray}
The background field  $ \rho_B $ is given by
\begin{equation}
 \rho_B=\sum\limits_{\kappa \in \{\uparrow, \downarrow\}}
 \langle|{\cal P}_{CS} \,c_\kappa^{+}(\vec{r}) \;
{\cal P}_{CS}^{+}c_\kappa(\vec{r})|\rangle \eqlabel{30096} \;.
\end{equation}
$ {\cal P}_{CS} $ is
constructed from $ P$ in (\ref{30080}) by the substitution
$ \vec{A} \rightarrow \vec{A}+\vec{a}_{CS} $. 
Because of this equation the Hartree graphs in the $ \nu=5/2 $
Chern-Simons theory are cancelled by the $ \rho_B $ couplings. 
Like in the $ \nu=1/2 $ system we are able to calculate
the Coulomb energy  of the path integral $Z_{5/2}^s $ in RPA.
Like in the case of the $ \nu=1/2 $ system
we obtain $ \mu_s=0 $ with the inclusion of the
spin constraint (\ref{30145}).
We now calculate the Coulomb energy of the path integral
(\ref{30150}), (\ref{30093}),
with the approximation
$ P_{CS}\approx  (1/\sqrt{2B_{\tiny \mbox{eff} }})\,
\Pi_{L,CS} $ in RPA. This approximation is also used in (\ref{30096}). 
For this we define the $ \nu=5/2 $ Coulomb potential by
$ V^{ee}_{5/2}(\vec{r})=e^2/(2B_{\mbox{\tiny eff}} r)$.   
With the help of the expressions 
\begin{eqnarray}
\mbox{EM}_{5/2} & = &  
 2\, (2\pi \tilde{\phi}) 
\frac{2\pi i }{q}V_{5/2}^{ee}(q)\,\Pi^d_{\sigma a^0}
\Pi^d_{a \sigma} \eqlabel{30340} \\
& & \nts{15} -(2\pi \tilde{\phi})^2 
\frac{2\pi}{q^2}V_{5/2}^{ee}(q)\,\Pi^d_{a^0 a^0}
\Pi^d_{\sigma a}\Pi^d_{a \sigma}
\nonumber \\
& & 
\nts{15}-(2\pi \tilde{\phi})^2 
\frac{2\pi}{q^2}V_{5/2}^{ee}(q)\,\left(\Pi^d_{a a}+
\frac{\mu_\uparrow+\mu_\downarrow}{2\pi}\right)
\Pi^d_{\sigma a^0}\Pi^d_{a^0 \sigma} \;,\nonumber \\
 \mbox{F}_{5/2}^{ee} & = &   -
\sum\limits_{\kappa \in \{\uparrow, \downarrow\}}\eqlabel{30320}    \\
 & &\nts{30}  \frac{1}{2(2\pi)^3}\int d^2k\, d^2q\,
k^2 q^2\,
V_{5/2}^{ee}(|\vec{k}-\vec{q}|)\;n_F(k,\mu_\kappa)\,n_F(q,\mu_\kappa)
\nonumber  
\end{eqnarray}
we get the RPA Coulomb energy by 
\begin{eqnarray} 
 \Omega^{ee}_{5/2}&  = & \mbox{F}_{5/2}^{ee}\eqlabel{30345} \\
& & +\frac{1}{2(2\pi)^3}\int 
 d\omega \,d^2q\; 
 \frac{1}{(1+\mbox{CS}^s(q,\omega))}\mbox{EM}_{5/2}(q,\omega)  \;.
 \nonumber
\end{eqnarray} 

\begin{figure}   
\centerline{\psfrag{y}{\turnbox{180}{\hspace{-2.0cm}\scriptsize Energy Density
 $\times (2 B_{\tiny \mbox{eff}})^2\;[e^2/r_s^7]$}}
 \psfrag{x}{\scriptsize $ 2 q^2/k_F^2 $}  
    \epsfig{file=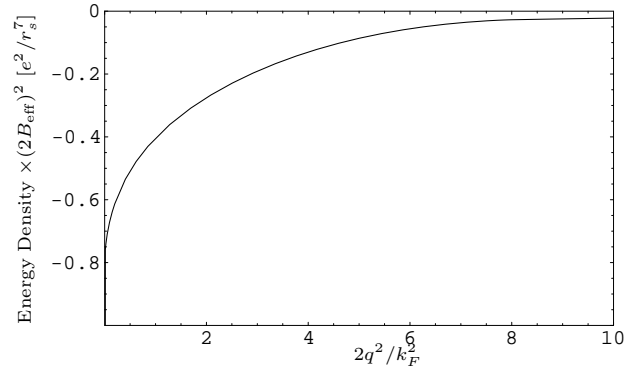,width=8cm}}
\caption{
  The contribution of the scaled momentum $ 2 q^2/k_F^2 $ to 
  $ (2 B_{\tiny \mbox{eff}})^2$  times  the Coulomb energy density
  for the spin polarized $ \nu=5/2 $ system. 
 $ (2 B_{\tiny \mbox{eff}})^2$  times  the energy density is shown
 in units of $e^2/r_s^7 $ where $ r_s $ is the
 effective electron distance $ \rho=1/(\pi r_s)^2 $ .
\label{param52}}
\end{figure}

The several  response functions $ \Pi $ in (\ref{30340}) are defined in
(\ref{30270}) of the appendix.
Now we have to fix the effective magnetic field
$ B_{\tiny \mbox{eff}} $ by perturbational methods. 
We will treat at first the case of a spin polarized ground state
(We have to insert $ \mu_\downarrow=0 $ and $
\Pi(q,\omega,\mu=0)=0$ in
(\ref{30340}) and (\ref{30320})). 
For fixing $ B_{\tiny \mbox{eff}} $ we have look for the main
contributions
to the momentum integrals of $ \Omega^{ee}_{5/2} $ in (\ref{30345}).
We will fix $ B_{\tiny \mbox{eff}} $ so that we get for these momentums the
same integrand as we would calculate the path integral (\ref{30150}),
(\ref{30093}) with the exact operator $P $ (\ref{30080}) in 
$L^{5/2}_{ee,l} $.  
Numerical integration of the two terms in  
$ \Omega_{5/2}^{ee} $ (\ref{30345}) yields results for $ \mbox{F}^{ee}_{5/2} $
which are about a factor
3 smaller compared to the second term.  
This is in contrast to the $ \nu=1/2 $ system.
When fixing $ 1/B_{\mbox{\tiny eff}}^2 $ in $V^{ee}_{5/2} $
there is no impact on the proportion
of the two terms.
Thus we use the second term in (\ref{30345}) for
fixing $ B_{\mbox{\tiny eff}} $. After performing the $ \omega $
integration in (\ref{30345}) the $ q$ integrand
of the second term has the largest contribution to the energy for
$ q \approx 0$. This is shown in figure \ref{param52}.
Thus the ring momentums  $ k \approx k_F $
are the most important wave vectors in the calculation of
the response
functions (\ref{30270}) in order to calculate
the ground state energy of the 
$ \nu=5/2 $ system. 
This is due to a $ n_F(\vec{k}+\vec{q})-n_F(\vec{k}) $ term 
as a result of integrating (\ref{30270}) over the ring frequencies.  
Thus we find the following equation to get the same  $ q\approx 0  $-integrand
in the second term in (\ref{30345}) 
using  the exact $ P_{CS} $ (\ref{30080}) and its  approximation
(\ref{30091}) 
in the path integral 
\begin{eqnarray}
& &  \langle
 u_{\vec{k}_F}|P_{CS}P^+_{CS}|u_{\vec{k}_F}\rangle\Big|_{\vec{a}=\vec{A}}
 \eqlabel{30390} \\
& & \hspace{1cm}  = \frac{1}{2B_{\mbox{\tiny eff}}} \langle
 u_{\vec{k}_F}|\Pi_{L,CS}\Pi^+_{L,CS}|u_{\vec{k}_F}\rangle\Big|_{\vec{a}=\vec{A}}=
 \frac{1}{2B_{\mbox{\tiny eff}}} k_F^2\;.\nonumber 
\end{eqnarray}
$ u_{\vec{k}_F} $ is the one particle function $ u_{\vec{k}_F}=1/\sqrt{F}
\exp[i \vec{k}_F \vec{r}] $ where
$ \vec{k}_F $ is a vector of length $ k_F=\sqrt{2m\mu} $.
Furthermore, we have to insert the mean-field condition
$ \vec{a}=\vec{A} $ in $P_{CS} $ of equation
(\ref{30390}). By using the definitions (\ref{30038}), (\ref{30080}) the left hand side of equation (\ref{30390}) is equal
to
\begin{eqnarray} 
& &  \nts{10} \langle
 u_{\vec{k}_F}|P_{CS}P^+_{CS}|u_{\vec{k}_F}\rangle\Big|_{\vec{a}=\vec{A}}
  \eqlabel{30391}\\
& &  \hspace{1cm}  =\lim\limits_{k \to 0 }\;
\langle u_{\vec{k}}|\,P\, P^+\,
|u_{\vec{k}}\rangle\Big|_{B_H= k^2 \, B_{\mbox{\tiny eff}}/k_F^2 }\nts{10}
 = 1 \;. \nonumber 
\end{eqnarray}
In this equation $ P$ is calculated
with the vector potential $ \vec{A}= B_H/2 \;(-y,x) $.
The magnetic field $ B_H $ is defined by $B_H= k^2 \, B_{\mbox{\tiny eff}}/k_F^2$.   
From the equations (\ref{30390}) and (\ref{30391}) we obtain 
$ B_{\mbox{\tiny eff}}=m \mu $.\\
In the case of
a spin unpolarized ground state we get
with the help of $\mu=\mu_\uparrow=\mu_\downarrow $
the approximation $ B_{\mbox{\tiny eff}}=m \mu $ 
due to the validity of the equations
(\ref{30390}), (\ref{30391}).
In the case of $ \mu_\uparrow \not=\mu_\downarrow $
we obtain  the equation
$(\mu_{\uparrow}+\mu_{\downarrow})/B_{\mbox{\tiny eff}}=2 $ to get the same
$ q\approx 0  $-integrand
in the second  term of equation (\ref{30345}) 
using  the exact $ P_{CS} $ (\ref{30080}) and its  approximation
(\ref{30091}) 
in the path integral (\ref{30150}), (\ref{30093}).  
If we compare this equation for $ \mu_\downarrow=0 $ with (\ref{30390}) and
(\ref{30391}) we get a factor two difference. This is because we used 
for the response functions $ \Pi(q,\omega,\mu=0)=0 $ in
equations (\ref{30390}) and
(\ref{30391}). This is only correct for $ |\omega| >0 $ (see e.g. $
\Pi_{a^0 a^0} $ in (\ref{30280})). One has to notice that
$ \Pi(q,\omega,0) $ is non-zero only on a non-measurable subspace of the
$ (\omega,q)$-integral of the second term of
$ \Omega^{ee}_{5/2} $ (\ref{30345}). Thus $ \Pi(q,\omega,0) $
is not relevant for the energy
calculation and $ B_{\mbox{\tiny eff}}=m \mu $ is a good
approximation for $ \mu_\downarrow \not=0 $.
The absolute value $ |\Pi(q,\omega,\mu)|$ of the response functions
(with any vertex functions)
are growing functions of
$ \mu$  for $ |\omega| > 0 $ and zero for $ \mu=0 $ and $ |\omega| > 0 $.
Therefore, for a determination of
$ B_{\mbox{\tiny eff}} $
we only consider that part of
the $ q\approx 0  $ reponse functions of the second term of
$ \Omega^{ee}_{5/2} $ (\ref{30345})
which contains the larger $ \mu $-value $ \mu_l $ of
$ \mu_\uparrow $, $ \mu_\downarrow $. The demand 
that this part of the $ q\approx 0  $-integrand is identical by 
using  in the path integral either  the exact $ P_{CS} $ (\ref{30080}) or its  approximation
(\ref{30091}), results in 
$ B_{\mbox{\tiny eff}}=m \mu_l $. Thus we get for $ B_{\mbox{\tiny eff}}
$ of the spin polarized
as well as the spin
unpolarized case the former values. \\[0.2cm]
In the following we calculate the compressibility, the effective mass
and the excitations of the
spin polarized $ \nu=2+1/\tilde{\phi} $ systems.
This is done by a restriction of the path integral 
(\ref{30150}), (\ref{30093}) to one spin component (and inserting
$ \mu_s=0 $).
We now use standard techniques \cite{hlr} to calculate the
compressibility, the effective mass and 
the excitations of the $ \nu=2+1/\tilde{\phi} $ systems in RPA.
To calculate the compressibility we have to pay attention to the
construction of
the path integral of the $ \nu=2+1/\tilde{\phi} $ systems.
It is then easy to see that in the above model of the
spin polarized $ \nu=2+1/\tilde{\phi} $ systems
the density-density correlation function is given by 
$ \langle |\rho^{5/2}(\vec{r}, l \epsilon)\rho^{5/2}(\vec{r}\,', l' \epsilon)|
\rangle $. In this expression $ \rho^{5/2}(\vec{r},l \epsilon)) $ is
given by the
quantum mechanical Heissenberg operator of the expression
$ {\cal P}_{CS} \,c_{\uparrow}^{+}(\vec{r}) {\cal P}_{CS}^{+}
c_{\uparrow}(\vec{r}) $.
The Fourier transformed density-density
correlation function
could be calculated by analytical continuation of the 
Fourier transformed sum of
$ \langle |\sigma_{l}(\vec{r})\sigma_{l'}(\vec{r}\,')|\rangle  $ and
$ (V^{ee})^{-1} (\vec{r}-\vec{r}\,') $. 
$ \langle |\sigma_{l}(\vec{r})\sigma_{l'}(\vec{r}\,')|\rangle  $ is the
$ \sigma \sigma $ correlation function of the path integral
(\ref{30150}), (\ref{30093}).
With the help
of the approximations (\ref{30430}) we obtain for
the (retarded) density-density correlation function \cite{mah1}
of the $ \nu=2+1/\tilde{\phi} $
system in the range $ \omega=0 $ and $ q \approx 0 $   
\begin{eqnarray}
& & \langle| T \rho^{5/2}(-q,0)\rho^{5/2}(q,0)|\rangle_{ret} \eqlabel{30500}  \\
& & \hspace{0.8cm} =\frac{\frac{4 m^3 \mu^2}
  { (2 B_{\tiny \mbox{eff}})^2}\left(1-\tilde{\phi}+\frac{\tilde{\phi}^2}{4}\right)}{2 \pi
  \left(\frac{4 m^3 \mu^2}{(2 B_{\tiny \mbox{eff}})^2}
      \,V^{ee}(q)\left(1-\tilde{\phi}+\frac{\tilde{\phi}^2}{4}\right)+
\left(1+\frac{\tilde{\phi}^2}{12}\right)\right)}\;. \nonumber 
\end{eqnarray}
Here $ T $ is the time ordering operator. From (\ref{30500}) we see that
$ \lim_{q \to 0}
\langle| T \rho^{5/2}(-q,0)\rho^{5/2}(q,0)|\rangle_{ret} $ is zero for
the $ \nu=5/2 $ system.
By considering higher order terms of the response functions (\ref{30280})
in the momentum $ q $ and $  \omega=0 $ we get for the $ \nu=5/2 $ system 
$ \langle| T \rho^{5/2}(-q,0)\rho^{5/2}(q,0)|\rangle_{ret}=m^2 q^2
/(3 \pi (2B_{\mbox{\tiny eff}})^2 (1+\tilde{\phi}^2/12))+O(q^4) $.
With the help of the retarded density-density correlation function
it is possible two calculate the compressibility $ K^{5/2} $
of the system by using the compressibility sum rule \cite{mah1}
$ K^{5/2}= \lim_{q\to 0}
(1/\rho^2) \langle| T \rho^{5/2}(-q,0)\rho^{5/2}(q,0)|\rangle_{ret}/
 (1-(2\pi) V^{ee}(q) \langle| T \rho^{5/2}(-q,0)\rho^{5/2}(q,0)|\rangle_{ret}
 ) $.
Due to this relation 
we obtain by the definition of $ B_{\tiny \mbox{eff}} $
that the compressibility is given by
$ K^{5/2}=m/(2 \pi \rho^2) (1-\tilde{\phi}+\tilde{\phi}^2/4)/
(1+\tilde{\phi}^2/12) $. 
Thus we see that the systems of filling fraction $ \nu=2+1/\tilde{\phi} $
($\tilde{\phi}\not=2$) should be compressible. 
Furthermore we get the same RPA compressibility as HLR calculated for the
$ \nu=1/2 $ system except for a factor
$ (1-\tilde{\phi}+\tilde{\phi}^2/4)$. In the case of filling fraction
$ \nu=5/2 $ this is no longer valid.
Due to our calculation the $ \nu=5/2 $ system is incompressible. 
Finally, if we take into account
all terms of $ P$ (\ref{30080}) 
in the RPA calculation of the compressibility, 
we get the same expression as (\ref{30500}) for the
$ \nu=2+1/\tilde{\phi}  $ systems.
This is due to  the definition of $ B_{\mbox{\tiny eff}} $ (\ref{30390}). \\
Next we calculate the effective mass
$ m^*=m \, (1- \partial/(\partial \omega)  \Sigma(k_F,0))/(
1+m/k_F \partial/(\partial k)  \Sigma(k_F,0)) $ in RPA. Here
$ \Sigma(k,\omega) $ is the self energy of the fermions of the path integral
(\ref{30150}), (\ref{30093}).
Due to the same singular structure of the transversal
($ a,a$) propagator we obtain for the $2+1/\tilde{\phi} $ system with the help of
(\ref{30430}) a singular effective mass which is different by 
a factor
$ 1/(1-\tilde{\phi}+\tilde{\phi}^2/4) $ 
from the (diverging) effective mass of the spin polarized $ \nu=1/2 $ system
calculated by HLR.
The expression for the compressibility as well as the effective mass suggests
that systems of filling fraction $ \nu=2+1/\tilde{\phi} $
behave similar to the  $ \nu=1/2 $ system except $\tilde{\phi}=2  $ . 
This is in agreement  with numerical calculations by Morf and d'Ambrumenil
\cite{mo1}. \\
Like in the $ \nu=1/2 $ system one can also derive
the cyclotron excitations for the $ \nu=2+1/\tilde{\phi} $
systems by a calculation of the 
density-density correlation function. In this calculation one has to take into
account the response functions in the limit (\ref{30510}).      

\section{ The results of the energy calculation \label{k5}}

\begin{figure}
 \centerline{   
     \epsfig{file=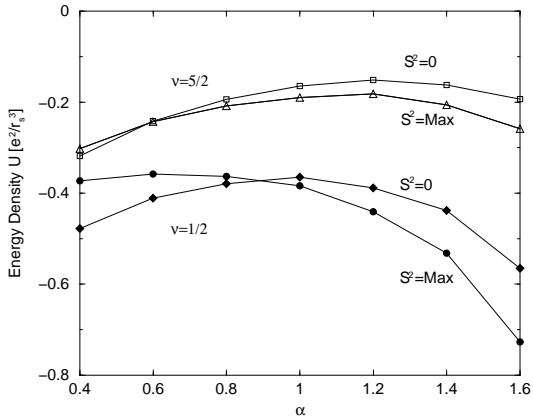,width=7cm}
 }
 \caption{The Coulomb energy of the ground state of the 
   $\nu=1/2$ and the $ \nu=5/2 $ system 
 including the spin constraints $S^2=0 $
 and $S^2=\mbox{Max}$. The Coulomb energy is given in units of
 $e^2/r_s^3 $ where $ r_s $ is the
 effective electron distance $ \rho=1/(\pi r_s)^2 $. \label{bild1}}
 \end{figure}

We have calculated the integrals in (\ref{30370}), (\ref{30345}) 
using numerical methods. The results are shown in figure \ref{bild1}.
This figure shows the Coulomb energy as a function of the strength 
$ \alpha $ of 
the Coulomb interaction function $V^{ee}(k)=e^2 /k^{\alpha} $. 
It contains the calculation of the spin polarized system  
, i.e. $ \mu_{\uparrow}=2\pi \rho/m$, $\mu_{\downarrow}=0 $, as well as 
the spin unpolarized system, i.e. 
$ \mu_{\uparrow}=\mu_{\downarrow}=\pi \rho/m $, for the $ \nu=1/2 $ and 
the $ \nu=5/2 $ system. 
The graph reveals a transition from a spin singlet to a spin polarized 
ground state at $ \alpha \approx 0.9 $ for the $ \nu=1/2 $ system. 
In the $ \nu=5/2 $ system the transition is at $ \alpha \approx 0.6 $. 
This coincides  qualitatively with the numerical result of Jain et al. 
\cite{be1}, which has be obtained by exact diagonalization methods 
on a sphere. Since Jain et al. employed the Haldanes pseudo 
potentials  $ V_L $, where the strength of the Coulomb interaction 
is adjusted on the basis of the ratio of $V_0 $ to $ V_1 $     
we can not quantitatively compare the two results \cite{ha1}.  
That is, Jain et al. modify the strength of the  
pseudo potentials with the help of $ \alpha'$ in 
$ [\alpha' V_1,V_1,V_2,...] $.
For the Coulomb interacting $\nu=1/2 $ system the result is 
$\alpha'=2.0 $. For the Coulomb interacting $\nu=5/2 $ system
one gets $ \alpha'=1.45 $.
Jain et al. obtained a transition from a spin singlet to a spin polarized    
ground state for the $\nu=1/2 $ system at $\alpha'=1.1 $. 
They obtained a good overlap of the ground state wave function with a spin polarized
Chern-Simons wave function for $ \alpha'>1.1 $.
The same transition was observed at $\alpha'=1.2 $ for the $\nu=5/2 $ system.
Because  $\alpha' $ is dominated by the strength of 
the interaction of neighbouring electrons, 
the transition at $\alpha<1 $ showing in figure (\ref{bild1}) 
agrees with the results of Jain et al..
But comparing the value $ \alpha $ at the transition 
of the $ \nu=1/2 $ system with the $ \nu=5/2 $ system,
$ \alpha $ of the $ \nu=5/2 $ system is probably too small.
Furthermore, the slope of the two curves in figure \ref{bild1} of spin $S^2=\mbox{Max} $ 
($ S^2=|S_z|(|S_z|+1)$ with $ S_z=\rho F /2 $)   
and $S^2=0 $ of the $ \nu=5/2 $ system differs only slightly. 
Thus the $ \nu=5/2 $ system compared to the $ \nu=1/2 $ system can change 
the alpha-value of the transition very easily due to the   
higher orders of perturbation theory. \\
Morf shows in \cite{mo2} using similar  
numerical diagonalization methods like Jain et al. but  
considering a larger number 
of electrons that the ground state of the 
$ \nu=5/2 $ system is not a Chern-Simons 
wave function for $ \alpha=1 $ (Coulomb potential). This ground state wave function shows 
a large overlap with a spin polarized incompressible paired wave function.  
For smaller $ \alpha' $, respectively $ \alpha $, Morf could not 
identify the ground state wave function.
Recently Rezayi and Haldane \cite{rez1} could identify this
ground state as a compressible stripped phase by using similar numerical
diagonalization methods like Morf. 
For larger $ \alpha' $, respectively $ \alpha $, Morf \cite{mo2} obtains 
for the ground state 
wave function a large overlap with a spin polarized 
Chern-Simons wave function.
Based on this observation he interpreted the experiments of Eisenstein
et al.\cite{ei1}, mentioned in the introduction, that caused on the
tilted magnetic field the effective interaction between two neighbouring
electrons is increased.
Thus the ground state wave function of the system changes to 
a spin polarized Chern-Simons wave function.
At last we mention that Rezayi and Haldane get in \cite{rez1}
a transition of a compressible stripped phase
via an incompressible paired quantum Hall state to a Chern-Simons ground
state near $ \alpha'=1.45 $ even though they considered the pure spin polarized
$ \nu=5/2 $ system. This shows that the transition is an
interaction effect (not caused by the spin degree of freedom).
\\
Comparing for $ \alpha=1 $ the Coulomb energies
$  \Omega_{1/2}^{ee} $, $  \Omega_{5/2}^{ee} $ of the 
ground state (spin polarized)
with the Coulomb energies
$ U^{N}_{1/2} $, $ U^{N}_{5/2} $ of the
numerical calculation method \cite{pa1} we obtain
\begin{eqnarray} 
  \Omega_{1/2}^{ee}=-0.13 \, e^2 \rho^{\frac{3}{2}} &, &  U_{1/2}^{N}=-0.10 \, e^2 \rho^{\frac{3}{2}} \;, \eqlabel{30410} \\
\Omega_{5/2}^{ee}=-0.067 \, e^2 \rho^{\frac{3}{2}}   
&, & 
U_{5/2}^{N}=-0.088 \, e^2 \rho^{\frac{3}{2}} \;. \eqlabel{30415}
\end{eqnarray} 

\begin{figure}
 \centerline{\psfrag{RRR}{
     \smash{\raisebox{-0.1cm}{\scriptsize $\nts{8} S_z/(\rho F) $}}}  
  \epsfig{file=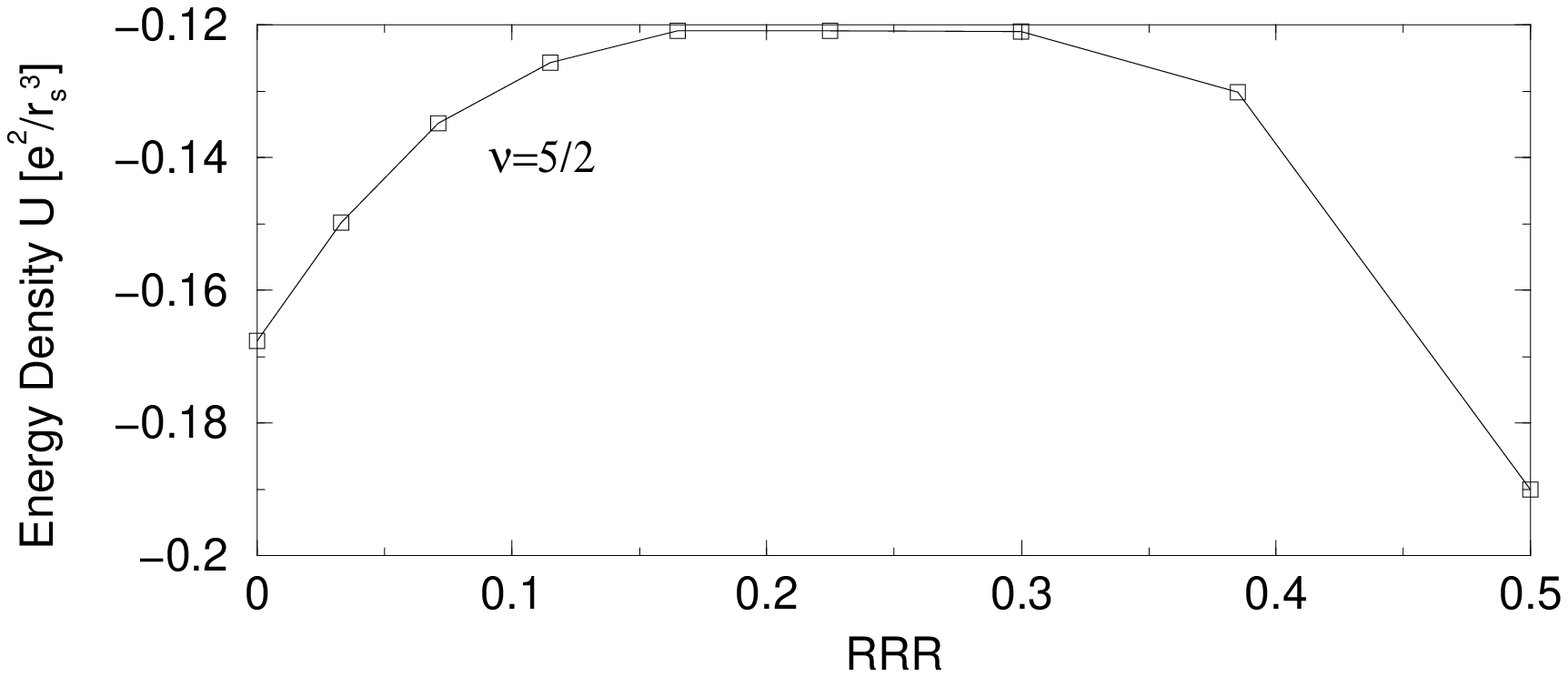,width=6cm}
   }\mbox{}\\[-0.2cm] 
 \centerline{\psfrag{RRR}{\smash{\raisebox{-0.1cm}{\scriptsize $\nts{8} S_z/(\rho F) $}}}
\epsfig{file=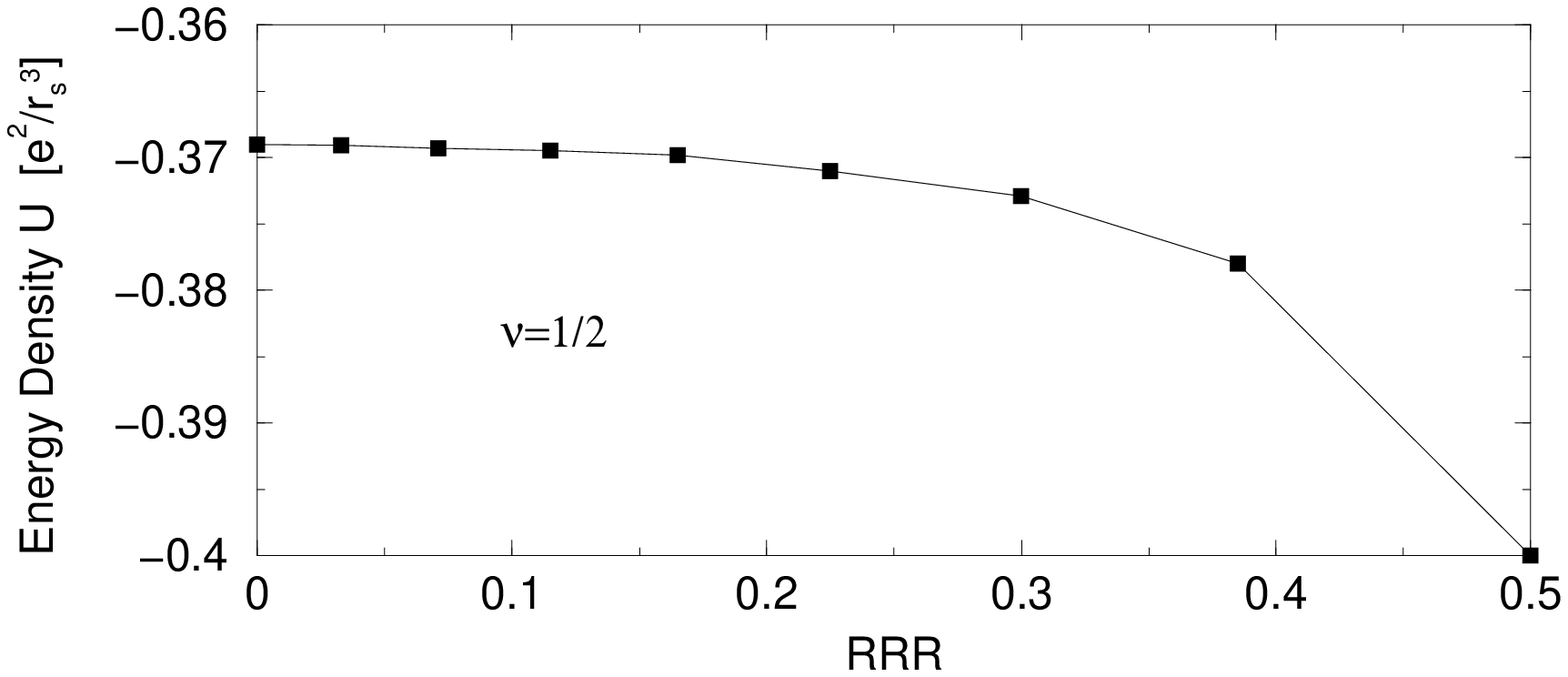,width=6cm}
  }
\\
 \caption{The Coulomb energy of the $\nu=1/2
 $ and the $\nu=5/2 $ system
 as a function of $ S_z/(\rho F) $. \label{bild2}
 The Coulomb energy is given in units of $e^2/r_s^3 $ where $ r_s $ is the
 effective electron distance $ \rho=1/(\pi r_s)^2 $.
 }
\end{figure}

Figure \ref{bild2} shows the Coulomb energy as a function of the variable
$ S_z/(\rho F) $ 
for $ \alpha=1 $
of the $\nu=1/2 $ and  
the $\nu=5/2 $ system.
$\mu_{\uparrow} $, $\mu_{\downarrow} $ are given 
by $ \mu_{\downarrow}=(1-2 S_z/(\rho F))/(1+2 S_z/(\rho F))\mu_{\uparrow}=
\pi (1-2 S_z/(\rho F))\rho/ m $.
The total spin $S^2 $
is determined by
$ S^2 =|S_z| (|S_z|+1) $  . 
The figure shows that the
spin polarized ground state, i.e. $ S^2=\mbox{Max} $, 
has a minimal Coulomb energy of all $ S^2 $ ground states.
This is in contradiction with recent experimental obeservations \cite{kuk1}
which shows that the $ \nu=1/2 $ ground state is not fully polarized.
Unfortunately these experiments do not show the degree of polarization of
the ground state. 

\section{Conclusion \label{k6}}
We showed that due to a wrong operator ordering in the path integral
of the spin polarized
$ \nu=1/2 $ system of HLR  we got in \cite{di1} an IR divergence in the
ground state magnetic energy. Therefore we used the well-known
Chern-Simons partition function in an operator formulation to get a
Chern-Simons path integral with a correct operator ordering. We calculated the
energy of this path integral in RPA to show that this
path integral has a UV infinite magnetic energy. This UV infinity can
be corrected by considering the complete set of Hartree-Fock Feynman graphs
of the Chern-Simons Hamiltonian. When calculating the
energy of the maximal divergent graphs together with the Hartree-Fock graphs
we obtained a well behaved finite magnetic energy.
The Coulomb energy is the same
as in \cite{di1}. \\
Next we developed a formalism to calculate
the Coulomb ground state energy of the $ \nu=1/2 $ system subjected to a  spin $ S^2 $
constraint and calculated the Coulomb energy as a function of the interaction
strength $ \sim e^2/k^\alpha$. This was done for a spin polarized ground state
as well as for a spin unpolarized ground state.
Furthermore, we formulated a Chern-Simons theory of the
$ \nu=2+1/\tilde{\phi} $ systems by transforming the interaction
operator of the electrons 
from the second to the lowest Landau level getting a
$ \nu=1/\tilde{\phi} $ Chern-Simons
theory with a modified interaction operator. We used this
Chern-Simons theory to calculate
the compressibility, the
effective mass and the excitations of the spin polarized
$ \nu=2+1/\tilde{\phi} $ systems in RPA. We see from our calculations that
$ \nu=5/2 $ is a special system within all $ \nu=2+1/\tilde{\phi} $
systems. We get the same compressibility for the $ \nu=2+1/\tilde{\phi} $
systems as HLR calculated for the $ \nu=1/2 $ system except for a factor
$ (1-\tilde{\phi}+\tilde{\phi}^2/4) $. 
Thus we get that the $ \nu=2+1/\tilde{\phi} $ systems are
compressible except for $ \tilde{\phi}\not=2 $. Due to our RPA calculation
the $ \nu=5/2 $ is incompressible. 
Next we calculated the effective mass of the $ \nu=2+1/\tilde{\phi} $
systems. 
For these systems we get a diverging effective mass like HLR got
for the $ \nu=1/2 $ system with a difference by a factor of 
$ 1/(1-\tilde{\phi}+\tilde{\phi}^2/4)$ ($\tilde{\phi} \not=2 $).
The effective mass of the $ \nu=5/2 $ system is the same as HLR calculated
for the $ \nu=1/2 $ system
without interaction between the electrons.
These calculations (compressibility and effective mass) are
in agreement with numerical calculations by Morf and d'Ambrumenil
\cite{mo1} which show that systems of filling fraction
$ \nu=2+1/\tilde{\phi} $ 
behave similar to the $ \nu=1/2 $ system
except for  $ \tilde{\phi} = 2 $. 
Like in the calculation of the cyclotron excitations of the $ \nu=1/2 $
system by HLR  we also obtain the cyclotron excitations for the
$ \nu=2+1/\tilde{\phi} $ systems in RPA. \\
Next we used the $ \nu=5/2 $
Chern-Simons theory to calculate the Coulomb energy
as a function of the interaction strength for the spin polarized
as well as the spin unpolarized ground state
in RPA.
We get for the $ \nu=1/2 $ as well as the $ \nu=5/2 $ system a transition
from a spin unpolarized ground state to a polarized ground state at
$ \alpha < 1 $. This agrees with calculations using
numerical diagonalization techniques. Furthermore, for $ \alpha =1$ we get
a good correspondence to the energies calculated by these numerical
methods. At last we
calculated the ground state Coulomb energies for the
$ \nu=1/2 $ and the $ \nu=5/2 $
system as a function of the total spin $ S^2 $ for $ \alpha=1$.  
We saw that for both systems the maximal spin polarized ground states
is actually in the Coulomb energy minimum of all $S^2 $ states which is not
in agreement with recent experimental observations \cite{kuk1}.\\ 
Summarizing, already on the level of the RPA,
we can see spin effects for both the $ \nu=1/2 $ and the $ \nu=5/2 $ 
systems which agree with numerical simulations. As mentioned earlier
it is supposed that the Coulomb $ \nu=5/2 $ Chern-Simons
ground state wave function ($\alpha=1$) is of a Pfaffian type.
Thus the goal still is 
to develop a theory of the Coulomb interacting $ \nu=5/2 $ system which has a
mean-field ground state of the paired wave type.   

\bigskip
We would like to thank M. Hellmund, K. Luig and W. Weller for many helpful 
discussions during the course of this work. 
Further we have to acknowledge the financial support by the Deutsche
Forschungsgemeinschaft, Graduiertenkolleg "Quantenfeldtheorie".

\begin{appendix}
\section{The response functions for the \boldmath $ \nu=1/2 $ \unboldmath
  and the \boldmath $ \nu=5/2 $ \unboldmath system \label{ka}} 
In this section we calculate the response functions for the $ \nu=1/2 $
and the $ \nu=5/2 $ system. These response functions were used in the
energy formulas (\ref{5105}), (\ref{30370}), (\ref{30345}). These response functions are defined by 
\begin{eqnarray}
 & & \Pi_{\sigma \sigma}(q,\omega,\mu)
    \eqlabel{30270} \\
 & & =\frac{1}{(2\pi)^3}
 \int d\Omega\, d^2k\;
 (\vec{k}+\vec{q})^2 \, \vec{k}^2 \,G(\vec{k}+\vec{q},\Omega+\omega)
 \,G(\vec{k},\Omega) \;, \nonumber \\
&&  \Pi_{\sigma a^0}(q,\omega,\mu) =  
 \Pi_{a^0 \sigma}(q,\omega,\mu)  \nonumber \\
&&  =\frac{1}{(2\pi)^3} \int d
 \Omega\, d^2k  
 \; (\vec{k}^2+\vec{k}\cdot\vec{q}) \;
 G(\vec{k}+\vec{q},\Omega+\omega)
 \,G(\vec{k},\Omega) \nonumber \;,\\
&& \Pi_{\sigma a}(q,\omega,\mu) =  -\Pi_{a\sigma}(q,\omega,\mu)
  \nonumber  \\ 
&& = -\frac{i}{(2\pi)^3} \int d\Omega\, d^2k\;   
 \frac{(\vec{k}\times\vec{q})^2}{m q} \; 
 G(\vec{k}+\vec{q},\Omega+\omega)
 \,G(\vec{k},\Omega) \;,\nonumber \\
&&  \Pi_{a^0 a^0}(q,\omega,\mu) \nonumber \\
&& =   \frac{1}{(2\pi)^3} \int d\Omega\, d^2k\;  
 G(\vec{k}+\vec{q},\Omega+\omega)
 \,G(k,\Omega) \;,\nonumber \\
&&  \Pi_{a a}(q,\omega,\mu) \nonumber \\
&&   =   \frac{1}{(2\pi)^3} \int d\Omega\,d^2k\;  
 \,\frac{(\vec{k}\times\vec{q})^2}{m^2 q^2} \;
 G(\vec{k}+\vec{q},\Omega+\omega)
 \,G(\vec{k},\Omega) \;.\nonumber 
\end{eqnarray}
G is the Green's function $ G(\vec{q},\omega)=-1/(i \omega-q^2/(2m)+\mu) $. 
After performing the integrals in (\ref{30270}) one gets with the
help of the substitution $u=\frac{q^2}{2m}$
\begin{eqnarray}
&& \Pi_{\sigma \sigma}^1(q,\omega,\mu) =   \frac{m^3}{4 \pi} 
\sum_{\sigma \in \{-1,1\}}\bigg[
+\frac{4}{30 u^3}
(u+i\sigma\omega)^5 \eqlabel{30280} \\
&& 
-\frac{1}{15u^3}\sqrt{(u+i\sigma\omega)^2-4u\mu}
\nonumber \\
& & \hspace{1.0cm} \times  
\left(2 (u+i\sigma\omega)^4+ 4(u+i\sigma\omega)^2u\mu+12\,u^2 \mu^2\right)
\bigg] \;,\nonumber \\
&& \Pi_{\sigma \sigma}^2(q,\omega,\mu)  =  \frac{m^3}{4 \pi} 
\sum_{\sigma \in \{-1,1\}}\left[
2\frac{(i\sigma\omega)}{3u^2}
\bigg\{(u+i\sigma\omega)^3
\right. \nonumber \\
& & \left.-\sqrt{(u+i\sigma\omega)^2-4u\mu} 
\left((u+i\sigma\omega)^2+ 2u\mu\right)\bigg\}-4\mu^2
\right] \;, \nonumber \\
&& \Pi_{\sigma a^0}^1(q,\omega,\mu) =  \frac{m^2}{4 \pi} 
\sum_{\sigma \in \{-1,1\}}\left[-\frac{1}{3 u^2}
(u+i\sigma\omega)^3 \right. \nonumber \\
& &\hspace{0.8cm} \left.+\frac{1}{3 u^2}\sqrt{(u+i\sigma\omega)^2-4u\mu} 
\left((u+i\sigma\omega)^2+2u\mu\right)
\right] \;,
\nonumber \\
&& \Pi_{\sigma a^0}^2(q,\omega,\mu) =   \frac{m^2}{4 \pi} 
\sum_{\sigma \in \{-1,1\}}\Bigg[\frac{(u+i\sigma\omega)}{u}\nonumber  \\
&& \hspace{1.3cm} \times \left((u+i\sigma\omega)
-\sqrt{(u+i\sigma
\omega)^2-4u\mu} \right)-2\mu
\Bigg]\;,
\nonumber \\
&& \Pi_{\sigma a}(q,\omega,\mu)  =     
-i\, m^{\frac{3}{2}}\sqrt{2u}\;\Pi_{aa}(q,\omega,\mu)
\;,\nonumber \\
&& \Pi_{aa}(q,\omega,\mu)  =    
\frac{1}{2\pi}\nts{2} \sum_{\sigma \in \{-1,1\}}\nts{2} \bigg[
\frac{1}{12 u^2}\bigg\{ \left(
u+i \sigma \omega \right)^3  \nonumber \\
&&\hspace*{2.5cm} -\sqrt{\left(\left(u+i \sigma \omega \right)^2-
4 u \mu\right)^3}
\bigg\}-\frac{\mu}{2}\bigg],
\nonumber \\
&&\Pi_{a^0 a^0}(q,\omega,\mu)  =     
\frac{m}{2\pi}\nts{3}
\sum_{\sigma \in \{-1,1\}}\nts{3} \left[\frac{1}{2u}
\sqrt{\left(u+i \sigma \omega \right)^2-4 u \mu}-\frac{1}{2} 
\right]\nts{2}.\nonumber   
\end{eqnarray}
$\Pi_{\sigma\sigma} $ and $\Pi_{\sigma a^0}$ are given by  
\begin{equation}\eqlabel{30290}
\Pi_{\sigma \sigma}=\Pi^1_{\sigma \sigma}+\Pi^2_{\sigma \sigma},
\quad, \qquad 
\Pi_{\sigma a^0}=\Pi^1_{\sigma a^0}+\Pi^2_{\sigma a^0} \;.
\end{equation}

For calculating the compressibility, the effective mass and the excitations
we have to continue analytically the response functions $ \Pi $ to imaginary
frequencies \cite{hlr}. In the range  $\omega\ll\frac{q^2}{2m} $ and $
q\ll k_F $ we get
{ \renewcommand{\arraystretch}{1.8}
\begin{equation}\eqlabel{30430}
\begin{array}{c c}
{\displaystyle \Pi_{\sigma \sigma}(q,-i\omega,\mu)=-\frac{4\,m^3\,\mu^2}{2
    \pi},}&{\displaystyle \nts{2}\Pi_{\sigma a^0}(q,-i\omega,\mu)=
  -\frac{2\,m^2\,\mu}{2\pi},} \\
{\displaystyle \Pi_{\sigma a}(q,-i\omega,\mu)=
 i \, q \frac{ \,m \,\mu}{2\pi} ,} &
{\displaystyle\Pi_{a^0 a^0}(q,-i\omega,\mu)=-\frac{m}{2\pi} ,} \\
\multicolumn{2}{c}{\displaystyle \Pi_{a a}(q,-i\omega,\mu)=\frac{q^2}{24\pi
    m}-\frac{\mu}{2\pi}-i\sqrt{2m\mu}\frac{\omega}{2\pi q}.}
\end{array} 
\end{equation}
}
By multiplying every value in
(\ref{30430}) by $ \omega/(q k_F) $, 
one can calculate the asymptotic of the next term in
the expansion of the response
functions.
Comparing $ \Pi_{a a} $ in (\ref{30430}) with the corresponding
term of HLR \cite{hlr}, the result differs by a factor two in the
first
term of $ \Pi_{aa}$. This difference is also observed in the case of the
$ \nu=1/2 $ Chern-Simons system with impurities \cite{di2}. \\
Furthermore, we need the analytical continued response functions in the range
$\frac{q^2}{2m}\ll\omega $ and $ q \ll k_F $
for a calculation of the cyclotron excitations . In this range we
get
{\renewcommand{\arraystretch}{2.0}
\begin{equation}\eqlabel{30510} 
\begin{array}{c c}
{\displaystyle \Pi_{\sigma \sigma}(q,-i\omega,\mu)=
  O\left(\frac{q^2}{m \omega}\right)\nts{1},} &{\nts{4} 
  \displaystyle \Pi_{\sigma a^0}(q,-i\omega,\mu)=
 O\left(\frac{q^2}{m \omega}\right)\nts{1},} \\
{\displaystyle \Pi_{\sigma a}(q,-i\omega,\mu)=
  O\left(\frac{q^2}{m \omega}\right),} &
{\displaystyle\Pi_{a^0 a^0}(q,-i\omega,\mu)=
  \frac{\mu}{2\pi}\frac{q^2}{\omega^2} ,} \\
\multicolumn{2}{c}{\displaystyle \Pi_{a a}(q,-i\omega,\mu)+\frac{\mu}{2\pi}=\frac{\mu}{2\pi}\;.}
\end{array} 
\end{equation}
}

\end{appendix}

\end{multicols}
\end{document}